\documentstyle [12pt] {article}
\title{ Crystal-based Approach to Beam Collimation in RHIC and SNS }
\author{\underbar{V.M. Biryukov}
\\ {\em IHEP Protvino, RU-142284, Russia }
\\ {\small Electronic address:  biryukov@mx.ihep.su  }
\\ N. Catalan-Lasheras, A. Drees, N. Malitsky,
D. Trbojevic
\\ {\em BNL, Upton, 11973 NY } }
\date{Presented at ICAP, Darmstadt, 11-14 September 2000}
\begin{document}

\maketitle

\begin{abstract}
Bent crystal serving as a scraper of the beam collimation system
can channel halo particles directly into the absorber.
By means of computer simulations, we analyse the capabilities of crystal
technique for the beam cleaning process.
Two applications are considered:
the crystal collimator now being installed into RHIC for cleaning
of the fully stripped gold ions, and a similar system being developed for
the Accumulator Ring of the Spallation Neutron Source.
\end{abstract}

\section{
Introduction }

Classic two-stage collimation systems for loss localisation
in accelerators typically use a small scattering target as a primary
element, whereas the secondary element is a bulk absorber \cite{jeann}.
Normally in colliders and storage rings the halo diffusion is pretty slow,
therefore the first touch of a halo particle with the aperture-restricting
collimator is rather a glancing touch, with the impact parameter on the
order of micron. Such a near-surface particle is easiely scattered out
of the collimator material.
The role of the primary element is to give a substantial angular kick
to incoming halo particles in order to increase the impact parameter
of the particles on the secondary element placed in some position,
optimised transversally and longitudinally for better interception.

Naturally, an amorphous target scatters particles in all possible directions.
Ideally, one would prefer a {\em "smart target"} that kicks all particles in
only one direction: for instance, only in radial plane, only outward,
and only into the preferred angular range corresponding to the center of
absorber (to exclude escapes).
Bent crystal is the first idea for such a smart target:
it traps particles as much as possible and convenes them into desired
direction. In physics language, we replace the scattering on single atoms
of amorphous target by the coherent scattering on atomic planes of
aligned monocrystal.

If the crystal channeling efficiency would be 100\%,
the crystal would serve as a kicker putting all the beam particles
deeply onto the collimator for safe absorption.
A real crystal is not 100\% efficient, therefore only part of the
beam goes into safe place, whereas the rest of the particles are
scattered and then handled in traditional way by collimators.

Several issues may be important for real application
of this technique for beam cleaning:
channeling efficiency, withstanding a high beam intensity,
long lifetime of crystal.
Recently the crystal team of IHEP has demonstrated the
following milestones.

\begin{itemize}
\item
The channeling crystal withstand the intensity of 2$\times$10$^{14}$
70-GeV proton hits per spill of $\sim$0.5 s duration \cite{1}.
This corresponds to averaged (over days) intensity of 2$\times$10$^{13}$
proton/s at crystal, already higher than the expected 0.1\% beam loss
in the SNS Accumulator Ring (=1.2$\times$10$^{13}$ p/s).
\newline
In a special pulse-mode test, one of the crystals was exposed to even
more extreme dump, $\sim$10$^{14}$ proton hits in short 50-ms pulse
with repetition rate once per 9.6 s.
The later external-beam test has shown that this crystal retained
normal channeling properties \cite{1}.
\item
Several crystals have been in exploitation at high intensity
(order of 5$\cdot$10$^{11}$ extracted protons in every spill)
for 1-2 months \cite{1}.
As a result, the integrated dose of proton hits at the crystal
has been about 10$^{20}$/cm$^2$
After this irradiation to 10$^{20}$/cm$^2$
the crystal extraction efficiency remained the same as measured in
the beginning of physics run and at the end.
Though high it is, the achieved integral irradiation is still below
the world highest results obtained in BNL \cite{bnl}
and CERN \cite{cern},
(4--5)$\times$10$^{20}$ proton/cm$^2$. The CERN experience showed that
at the achieved threshold of
5$\times$10$^{20}$ p/cm$^2$ at 450 GeV
the crystal lost 30\% of its deflection  efficiency.
At 450 GeV, crystal is sensitive to lattice misalignment of $\sim$5 $\mu$rad,
while at the SNS energy it is tolerant to $\sim$100 $\mu$rad, therefore
crystal may withstand higher doses at SNS, $\sim$10$^{21}$ p/cm$^2$.
One of the IHEP crystals did extract 70-GeV protons over 10 years without
replacement!
\item
The experimentally demonstrated figures of crystal deflection efficiency
are 65\% (IHEP, 70 GeV slow extraction \cite{1})
and 60\% (CERN, 450 GeV external beam deflection\cite{cern}),
here efficiency is the ratio of the deflected beam
to all incident beam.
It's been earlier demonstrated in IHEP experiment
that crystal scraper (channeling then with 50\% efficiency)
has reduced two-fold the
radiation levels downstream in the machine \cite{pac99}.
\end{itemize}

This year,
in the framework of the project to design and test a collimation system
prototype using bent channeling crystal for cleaning of the RHIC heavy ion
beam halo, a silicon (110) single crystal is being installed into the vacuum
chamber of one of the RHIC rings with proposition of being a primary element
situated upstream of the traditional heavy "amorphous" collimator \cite{rhic}.

The activities on crystal collimation concept have been largely supported
by simulations of the impact that the improvement in scraping efficiency
would have on the radiation backgrounds in the machines.
Making use of computer code CATCH \cite{catch} for crystal simulations,
earlier extensively verified in CERN, Fermilab, and IHEP experiments
in 1992-2000 \cite{fnal}, it's been shown in
Tevatron scraping simulations that crystal scraper
reduces accelerator-related background in CDF and D0
experiments by a factor of 10 \cite{tev}.

For the application in the SNS one can take a 100\%-safe approach:  using a
1-mm bent Si crystal together (just in front) with normal primary scraper
(like 5 mm Pt or W) of the collimation system, in the same location.  First,
this ensures that if crystal is not channeling then the system works as
usual. When crystal is brought to channeling condition, it works just as a
{\em ``crystalline edge''} for the primary target, deflecting (``shaving
off'') at $\sim$5  mrad as much  halo as possible. In this way, the
deflected particles will also be scattered then in the Pt target,
therefore the actual resulting deflection will be rather in some range;
this may even have an additional positive effect, spreading the beam load
on the absorber over some larger area. The crystal scraper protects the
edge of absorber from heat and radiation load.

The above said is a strong motivation to pursue the experimental tests
of crystal-assisted collimation of beams in order to
evaluate the potential benefits for the beam collimation systems.
The experimental studies on beam collimation in 1 GeV range are under
progress in the main ring of the IHEP U-70 accelerator, performed jointly
by the collaboration of BNL/SNS and IHEP during 2000 \cite{sns}.
The following stage of the
experimental program will make use of crystalline scrapers
dedicated to 1 GeV studies.

\section{ Crystal Scraper }

As the crystal can be aligned to the beam envelope, then to first
approximation all the incoming halo particles are parallel to the
crystal planes thus ensuring a high efficiency of channeling.
How parallel they are, depends on a beam growth rate.

In case of RHIC gold ion beam, the intrabeam scattering provokes a pretty
rapid growth of the beam. On the other hand, the location chosen for the
bent crystal has a huge beta function, in excess of 1000 m.
The earlier performed modelling of beam diffusion in RHIC \cite{rhic}
has provided us with a sample of halo particles that have a divergence
at the incidence on the crystal of a few microradian, well within the
critical angle of channeling in silicon,
$\theta_c$= $\pm$10 $\mu$rad at the RHIC energy of 250 GeV per unit charge.

The crystal situated $\sim$8 m upstream of a massive collimator is bending
the incoming particles by 0.5 mrad. As computer simulations show,
the efficiency of bending at 0.5 mrad for the considered sample of particles
is over 70\%. If we include all the particles bent at least 0.1 mrad
(they are still well intercepted by collimator), the bent fraction amounts
to 78\%.

For the SNS, the beam loss scenario is pretty unknown yet.
Therefore, we study the effect of crystal scraper \cite{pacs}
as a function of
divergence of the particles intercepted by a crystal.
Figure 1 shows the bending effect when the incident particles are well
within the angular acceptance of the silicon (110) crystal planes,
2$\theta_c$=0.25 mrad at the SNS kinetic energy of 1.3 GeV.

\begin{figure}[htb]
\begin{center}
\setlength{\unitlength}{0.9mm}
\begin{picture}(120,100)(-30,-6)
\thicklines
\linethickness{.5mm}
\put(    -20. ,1.8)  {\line(1,0){5}}
\put(    -15. ,2.4)  {\line(1,0){5}}
\put(    -10. ,6.0)  {\line(0,-1){3.6}}
\put(    -10. ,6.0)  {\line(1,0){5}}
\put(     -5. ,6.8)  {\line(1,0){5}}
\put(      0.,11.4)  {\line(0,-1){4.6}}
\put(      0.,11.4)  {\line(1,0){5}}
\put(      5.,11.6)  {\line(1,0){5}}
\put(     10. ,8.4)  {\line(0,1){3.2}}
\put(     10. ,8.4)  {\line(1,0){5}}
\put(     15. ,8.4)  {\line(1,0){5}}
\put(     20. ,8.8)  {\line(1,0){5}}
\put(     25. ,7.8)  {\line(1,0){5}}
\put(     30. ,6.6)  {\line(1,0){5}}
\put(     35. ,3.8)  {\line(1,0){5}}
\put(     35. ,3.8)  {\line(0,1){2.8}}
\put(     40. ,3.8)  {\line(0,1){2.4}}
\put(     40. ,6.2)  {\line(1,0){5}}
\put(     45. ,5.2)  {\line(1,0){5}}
\put(     50. ,104.8)  {\line(1,0){5}}
\put(     50. ,104.8)  {\line(0,-1){99.6}}
\put(     55. ,104.8)  {\line(0,-1){104.8}}

\linethickness{.4mm}
\put(-20,0) {\line(1,0){90}}
\put(-20,0) {\line(0,1){110}}
\put(-20,110) {\line(1,0){90}}
\put(70,0){\line(0,1){110}}
\multiput(-17.5,0)(10,0){9}{\line(0,1){2}}
\multiput(-20,20)(0,20){5}{\line(1,0){2}}
\multiput(-20,4)(0,4){25}{\line(1,0){1.}}
\multiput(70,20)(0,20){5}{\line(-1,0){2}}
\multiput(70,4)(0,4){25}{\line(-1,0){1.}}
\put(-7.5,-5){\makebox(1,1)[b]{-1}}
\put(2.5,-5){\makebox(1,1)[b]{0}}
\put(12.5,-5){\makebox(1,1)[b]{1}}
\put(22.5,-5){\makebox(1,1)[b]{2}}
\put(32.5,-5){\makebox(1,1)[b]{3}}
\put(42.5,-5){\makebox(1,1)[b]{4}}
\put(52.5,-5){\makebox(1,1)[b]{5}}
\put(-17,20){\makebox(1,.5)[l]{10}}
\put(-17,40){\makebox(1,.5)[l]{20}}
\put(-17,60){\makebox(1,.5)[l]{30}}
\put(-17,80){\makebox(1,.5)[l]{40}}
\put(-17,100){\makebox(1,.5)[l]{50}}

\put(-10,102){\large N(rel.)}
\put(30,-10){\large Angle (mrad)}

\end{picture}
\end{center}
\caption{
Angular distribution of 1.3 GeV protons downstream of a 0.5-mm single
silicon crystal scraper with 5-mrad bending,
for a parallel incident beam.
}
  \label{single}
\end{figure}
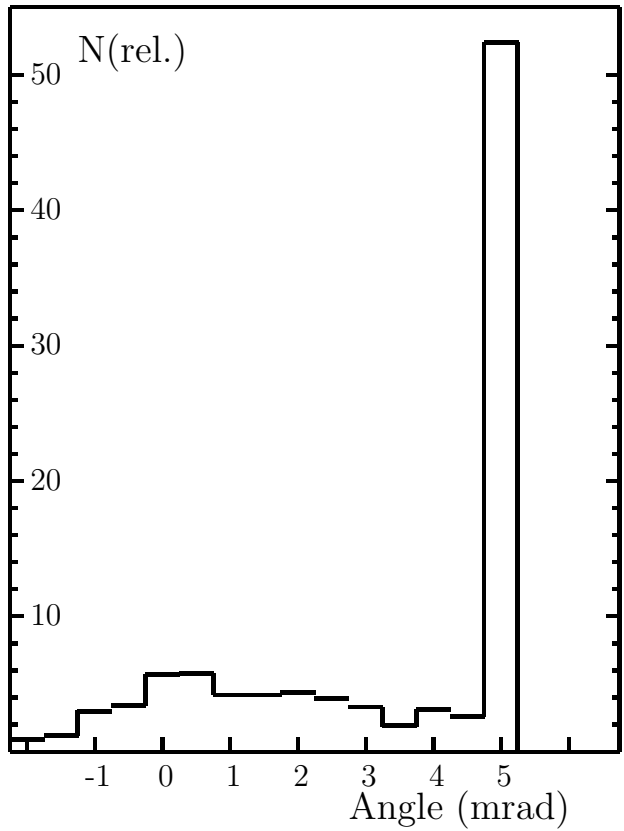

The particles that were not channeled in the crystal get some scattering,
nearely the same as in an amorphous matter.
Later, these unchanneled particles might be intercepted by crystal again
and channeled on this later encounter.
Multiple encounters of particles with channeling crystal essentially
improve the chances to enter the channeling mode in crystal \cite{bi91}.
In the experiments on "multipass" crystal extraction this happens
automatically, as the particles scattered in crystal go on circulating
in the ring and occasionally may encounter crystal again.

In some cases it's not that easy to have multiple encounters with a single
crystal. For instance, the SNS has a rapid-cycling Accumulator Ring,
60 Hz, with a beam lifetime about 1200 turns, and it's not obvious
that particles once scattered in a crystal will encounter it again.
Moreover, the behaviour of near-aperture particle can be quite complicated,
and the beam instability driving the halo can be quite fast.
In the case of RHIC, as a result of huge beta in the cleaning insertion,
the scattering in crystal increases the betatron amplitude by sort of
millimeters; this again makes it unobvious that the scattered particles,
after some circulation in the ring, will likely come to crystal again.

The alternative to "automatic" multiple encounters with a single crystal
scraper is to install several (two, three, or more) crystals in a row.
Then each of them must be aligned (or pre-aligned) in position and angle.
This would complicate the mechanics and procedures, but at the moment we
are interested only in potential physics benefits.

As the scattering of unchanneled particles in crystal increases their
divergence and hence affects the chances of channeling on a later
encounter with a crystal, one should further optimise the length of
a crystal to improve in overall probability of channeling as a result
of several encounters with a crystal.

For the same sample of gold particles in RHIC, we modelled their
successive encounters with a crystal.
For RHIC, the optimum crystal for 0.5 mrad bending is about the size
already chosen for the real prototype of crystal collimator now ready
for the beam tests. Figure 2 shows how the efficiency of crystal bending
increases with the number of particle encounters with a crystal
(4 mm long Si(110)).The first encounter provides the bending efficiency
of just under 80\%.
The bent fraction (we count here all particles deflected more than 0.1 mrad
toward the collimator) becomes order of 90\% after about 3rd encounter,
95\% after 7th, and over 97\% after about 10 encounters with 4-mm crystal.

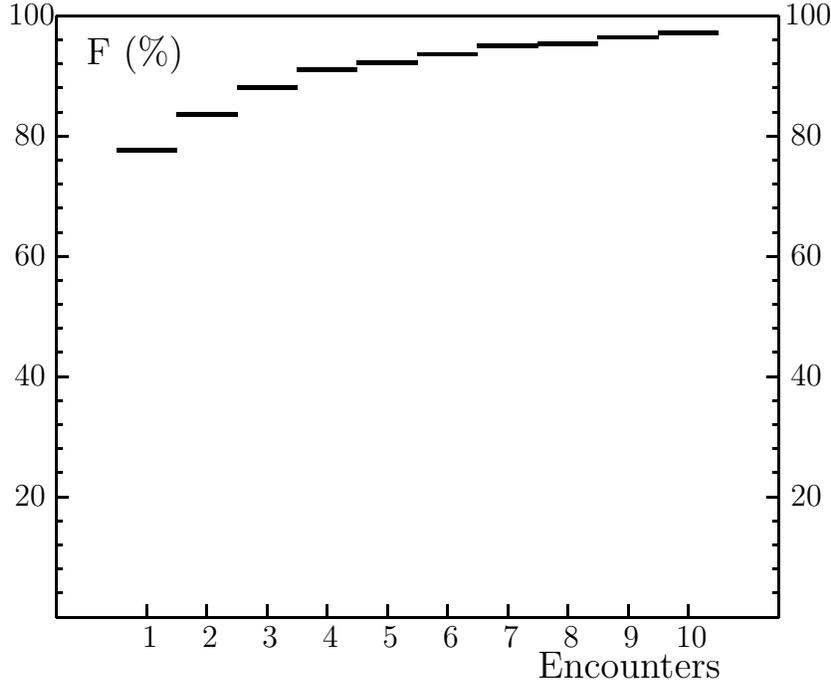
\begin{figure}[htb]
\begin{center}
\setlength{\unitlength}{0.8mm}
\begin{picture}(120,100)(-5,-6)
\thicklines
\linethickness{.5mm}
\put(      0. ,77.6)  {\line(1,0){10}}
\put(     10. ,83.6)  {\line(1,0){10}}
\put(     20. ,88.)  {\line(1,0){10}}
\put(     30. ,91.)  {\line(1,0){10}}
\put(     40. ,92.2)  {\line(1,0){10}}
\put(     50. ,93.6)  {\line(1,0){10}}
\put(     60. ,95.)  {\line(1,0){10}}
\put(     70. ,95.4)  {\line(1,0){10}}
\put(     80. ,96.4)  {\line(1,0){10}}
\put(     90. ,97.2)  {\line(1,0){10}}

\linethickness{.25mm}
\put(-10,0) {\line(1,0){120}}
\put(-10,0) {\line(0,1){100}}
\put(-10,100) {\line(1,0){120}}
\put(110,0){\line(0,1){100}}
\multiput(5,0)(10,0){10}{\line(0,1){2}}
\multiput(-10,20)(0,20){5}{\line(1,0){2}}
\multiput(-10,4)(0,4){25}{\line(1,0){1.}}
\multiput(110,20)(0,20){5}{\line(-1,0){2}}
\multiput(110,4)(0,4){25}{\line(-1,0){1.}}
\put(5,-5){\makebox(1,1)[b]{1}}
\put(15,-5){\makebox(1,1)[b]{2}}
\put(25,-5){\makebox(1,1)[b]{3}}
\put(35,-5){\makebox(1,1)[b]{4}}
\put(45,-5){\makebox(1,1)[b]{5}}
\put(55,-5){\makebox(1,1)[b]{6}}
\put(65,-5){\makebox(1,1)[b]{7}}
\put(75,-5){\makebox(1,1)[b]{8}}
\put(85,-5){\makebox(1,1)[b]{9}}
\put(95,-5){\makebox(1,1)[b]{10}}
\put(-17,20){\makebox(1,.5)[l]{20}}
\put(-17,40){\makebox(1,.5)[l]{40}}
\put(-17,60){\makebox(1,.5)[l]{60}}
\put(-17,80){\makebox(1,.5)[l]{80}}
\put(-18,100){\makebox(1,.5)[l]{100}}
\put(112,20){\makebox(1,.5)[l]{20}}
\put(112,40){\makebox(1,.5)[l]{40}}
\put(112,60){\makebox(1,.5)[l]{60}}
\put(112,80){\makebox(1,.5)[l]{80}}
\put(111,100){\makebox(1,.5)[l]{100}}

\put(-5,92){\large F (\%)}
\put(70,-10){\large Encounters}

\end{picture}
\end{center}
\caption{
The fraction of bent Au ions
as a function of the number of encounters with a crystal.
}
  \label{fnrhic}
\end{figure}

Although this figure may look impressive, 30-fold reduction of backgrounds
by crystal alone, it's not simple to realise technically.
Either one installs 3--10 crystals in a row in same location,
with a few-$\mu$m precision in relative radial position $x$
and a few-$\mu$rad precision in the angle $x'$;
which is possible but may be a substantial headache.
Or one installs a single crystal but takes care that particles
scattered of the crystal are not intercepted elsewhere
and likely will hit the crystal again on later turns.

Actually with every scattering in a 4-5 mm crystal the particles gain
$\sim$1 mm in betatron amplitude in this location.
This means that our crystal may have 2--3 encounters per particle,
after that the scattered particle is likely to be lost elsewhere.
Still, 2--3 encounters make a good improvement as shows Figure 2.

For the SNS the optimum for the crystal size should obviously be quite
smaller if we take into account increased scattering angles near 1 GeV.
Although a single crystal of $\sim$1 mm size (already in use in IHEP)
provides a similar efficiency for a parallel beam as in RHIC
case, a multi-crystal approach requires much shorter crystals.
Luckily, there is a technique of
graded composition Si$_{1-x}$Ge$_x$/Si strained layers
which form a bent crystal lattice of uniform curvature \cite{breese}
with any thickness along the beam direction from only 1 $\mu$m up to
millimeters, having a big cross-section (many centimeters) at the same time.
This technique limits the acievable bending angle at the SNS energy
to order of 2 mrad \cite{breese}.
We optimised the crystal size in that case and found the optimum crystal
to be about 70 $\mu$m along the beam, for the case of 2 mrad bending.

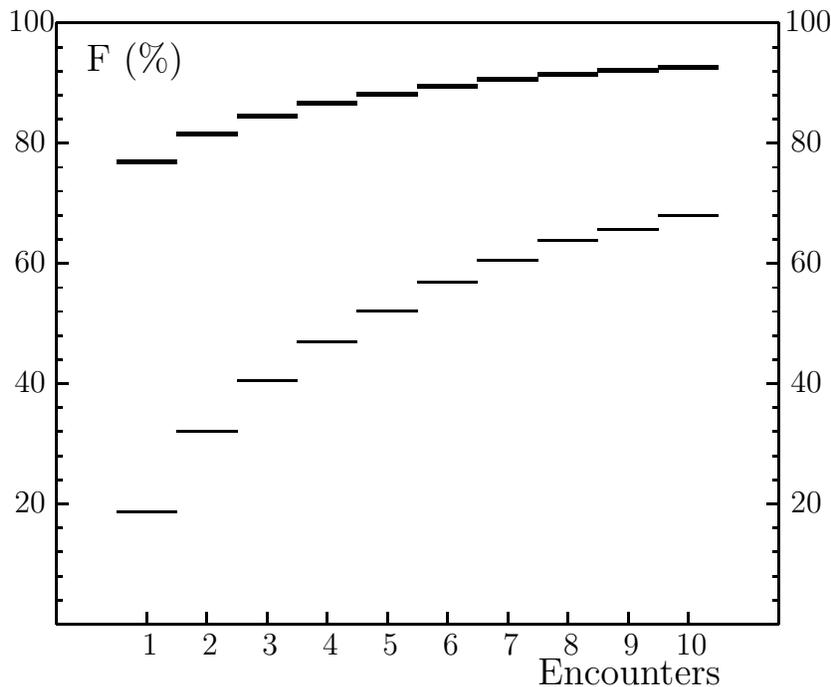
\begin{figure}[htb]
\begin{center}
\setlength{\unitlength}{0.8mm}
\begin{picture}(120,100)(-5,-6)
\thicklines
\linethickness{.5mm}
\put(      0. ,76.9)  {\line(1,0){10}}
\put(     10. ,81.6)  {\line(1,0){10}}
\put(     20. ,84.5)  {\line(1,0){10}}
\put(     30. ,86.6)  {\line(1,0){10}}
\put(     40. ,88.1)  {\line(1,0){10}}
\put(     50. ,89.4)  {\line(1,0){10}}
\put(     60. ,90.6)  {\line(1,0){10}}
\put(     70. ,91.4)  {\line(1,0){10}}
\put(     80. ,92.1)  {\line(1,0){10}}
\put(     90. ,92.6)  {\line(1,0){10}}

\linethickness{.25mm}
\put(      0. ,18.6)  {\line(1,0){10}}
\put(     10. ,32.0)  {\line(1,0){10}}
\put(     20. ,40.6)  {\line(1,0){10}}
\put(     30. ,47.0)  {\line(1,0){10}}
\put(     40. ,52.1)  {\line(1,0){10}}
\put(     50. ,56.9)  {\line(1,0){10}}
\put(     60. ,60.6)  {\line(1,0){10}}
\put(     70. ,63.8)  {\line(1,0){10}}
\put(     80. ,65.7)  {\line(1,0){10}}
\put(     90. ,68.)  {\line(1,0){10}}

\put(-10,0) {\line(1,0){120}}
\put(-10,0) {\line(0,1){100}}
\put(-10,100) {\line(1,0){120}}
\put(110,0){\line(0,1){100}}
\multiput(5,0)(10,0){10}{\line(0,1){2}}
\multiput(-10,20)(0,20){5}{\line(1,0){2}}
\multiput(-10,4)(0,4){25}{\line(1,0){1.}}
\multiput(110,20)(0,20){5}{\line(-1,0){2}}
\multiput(110,4)(0,4){25}{\line(-1,0){1.}}
\put(5,-5){\makebox(1,1)[b]{1}}
\put(15,-5){\makebox(1,1)[b]{2}}
\put(25,-5){\makebox(1,1)[b]{3}}
\put(35,-5){\makebox(1,1)[b]{4}}
\put(45,-5){\makebox(1,1)[b]{5}}
\put(55,-5){\makebox(1,1)[b]{6}}
\put(65,-5){\makebox(1,1)[b]{7}}
\put(75,-5){\makebox(1,1)[b]{8}}
\put(85,-5){\makebox(1,1)[b]{9}}
\put(95,-5){\makebox(1,1)[b]{10}}
\put(-17,20){\makebox(1,.5)[l]{20}}
\put(-17,40){\makebox(1,.5)[l]{40}}
\put(-17,60){\makebox(1,.5)[l]{60}}
\put(-17,80){\makebox(1,.5)[l]{80}}
\put(-18,100){\makebox(1,.5)[l]{100}}
\put(112,20){\makebox(1,.5)[l]{20}}
\put(112,40){\makebox(1,.5)[l]{40}}
\put(112,60){\makebox(1,.5)[l]{60}}
\put(112,80){\makebox(1,.5)[l]{80}}
\put(111,100){\makebox(1,.5)[l]{100}}

\put(-5,92){\large F (\%)}
\put(70,-10){\large Encounters}

\end{picture}
\end{center}
\caption{
The fraction of 1.3 GeV protons bent more than 1 mrad,
as seen downstream of a 10$\times$70$\mu$ multi-crystal
silicon scraper, as a function of the encountered number of crystals.
For a parallel incident beam (top, thick line),
and for the beam incident with $\pm$0.5-mrad divergence (bottom, thin line).
}
  \label{fnsns}
\end{figure}

Figure 3 shows how the bent fraction of 1.3 GeV protons
increases with the number of Si crystals encountered.
In this example we have first taken a parallel incident beam,
and showed in the Figure all the particles bent more than 1 mrad.
After the first encounter, 73\% of particles are found at 2 mrad,
in the channeled peak, and 77\% of particles are in the region
of $\ge$1 mrad.
The bent fraction comes over 90\% on about 6th encounter, and
totals near 93\% after 10 encounters with 70 $\mu$m crystal.

This simulation was repeated with a divergent incident beam
(1 mrad full width, flat distribution), with the results shown
in Figure 3 (thin line, bottom).
Very interesting feature is that -- whereas the efficiency of a single
encounter drops expectedly -- the multiple encounters boost the
bending effect essentially. The diverging particles, unchanneled first,
modify their angles with scattering and eventually fit the
crystallographic direction.
After 10 crystals, 56\% of the beam is found in the channeled
peak at 2 mrad.

\begin{figure}[htb]
\begin{center}
\setlength{\unitlength}{0.58mm}
\begin{picture}(220,100)(-15,-6)
\thicklines
\linethickness{.5mm}
\put( 0. ,92.6){\circle*{3}}
\put( 0. ,76.9){\circle{3}}
\put( 50. ,77.){\circle*{3}}
\put( 50. ,34.4){\circle{3}}
\put( 100. ,68.){\circle*{3}}
\put( 100. ,18.6){\circle{3}}
\put( 150. ,60.5){\circle*{3}}
\put( 150. ,14.){\circle{3}}
\put( 200. ,56.4){\circle*{3}}
\put( 200. ,12.3){\circle{3}}

\linethickness{.25mm}
\put(-10,0) {\line(1,0){220}}
\put(-10,0) {\line(0,1){100}}
\put(-10,100) {\line(1,0){220}}
\put(210,0){\line(0,1){100}}
\multiput(0,0)(50,0){5}{\line(0,1){3}}
\multiput(0,0)(10,0){20}{\line(0,1){1.5}}
\multiput(-10,20)(0,20){5}{\line(1,0){2}}
\multiput(-10,4)(0,4){25}{\line(1,0){1.}}
\multiput(210,20)(0,20){5}{\line(-1,0){2}}
\multiput(210,4)(0,4){25}{\line(-1,0){1.}}
\put(0,-9){\makebox(1,1)[b]{0}}
\put(49,-9){\makebox(1,1)[b]{0.5}}
\put(100,-9){\makebox(1,1)[b]{1}}
\put(149,-9){\makebox(1,1)[b]{1.5}}
\put(200,-9){\makebox(1,1)[b]{2}}
\put(-22,20){\makebox(1,.5)[l]{20}}
\put(-22,40){\makebox(1,.5)[l]{40}}
\put(-22,60){\makebox(1,.5)[l]{60}}
\put(-22,80){\makebox(1,.5)[l]{80}}
\put(-24,100){\makebox(1,.5)[l]{100}}
\put(212,20){\makebox(1,.5)[l]{20}}
\put(212,40){\makebox(1,.5)[l]{40}}
\put(212,60){\makebox(1,.5)[l]{60}}
\put(212,80){\makebox(1,.5)[l]{80}}
\put(211,100){\makebox(1,.5)[l]{100}}

\put(-5,105){\large F (\%)}
\put(120,-18){\large Divergence (mrad)}

\end{picture}
\end{center}
\caption{
The fraction of bent 1.3 GeV protons
as a function of divergence of the incident beam.
After one crystal (o) and after 10 crystal encounters ($\bullet$).
}
  \label{fdiv}
\end{figure}
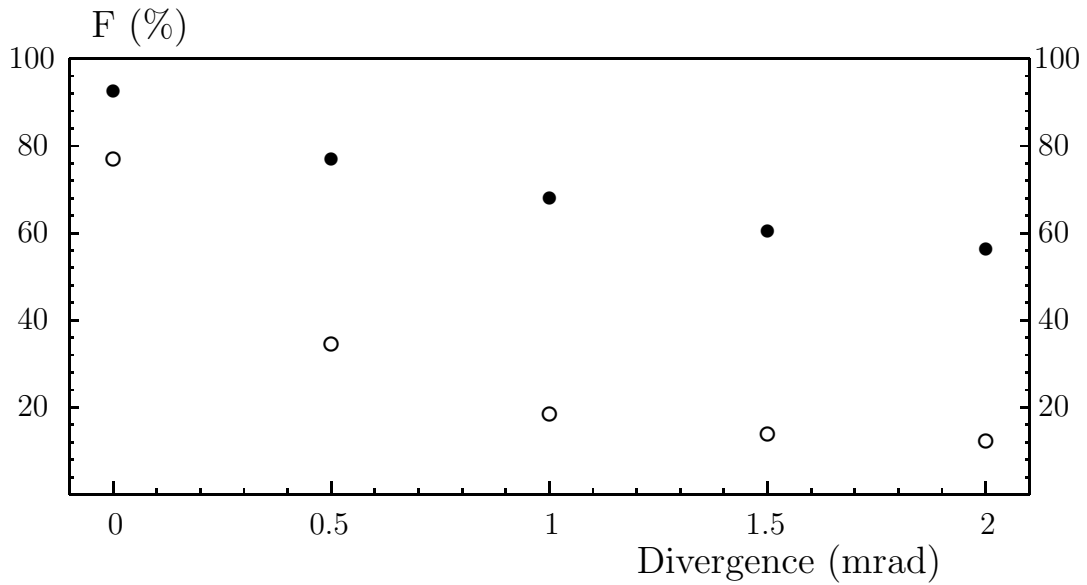

We studied further the effect of multi-crystal at several divergences.
As shows Figure 4, while a single crystal scraper drops the efficiency
when the divergence of incident particles goes in a range of milliradians,
the pack of ten 70-$\mu$m crystals is efficient through this range.

Essential technological difference of the SNS case from the RHIC one
is that all 10 wafers of Si$_{1-x}$Ge$_x$/Si
can be packed together in a single unit (about 10$\times$70$\mu$
=0.7 mm along the beam and some centimeters across)
before installing it into the accelerator.
The 10 wafers can be {\em pre-}aligned to each other when they are mounted
together into a single unit. This is easy in the SNS case because
the needed accuracy of alignment is low; as shows Figure 4,
if wafers are misaligned to the beam by sort of milliradian,
they still channel a substantial part of the beam.
This also means that the unit, 10-pack of crystals,
will have to be aligned to the beam envelope with same accuracy;
an accuracy of a fraction of milliradian might do the job.
Potentially, such a low demand for accuracy might even mean
that after installation of the crystal target into vicinity of the beam
one doesn't need anymore adjustment of its angle.

\section{Conclusion}

Our computer simulations of crystal channeling
show that 80-95\% of the incident halo particles
in the rings of RHIC and SNS can be channeled
directly into the absorber.
The collimation system then has to intercept the remaining few particles
only. This may give the factor of 5-20 improvement in the
collimation efficiency.

An approach with multi-crystal scraper can be useful also in the cases
where the same scraper must handle beams in a broad range of energies
(e.g., during acceleration). Then the first crystal can be thin,
dedicated to low-end of the energy range, while the next crystal is
optimised for high-end of the range.

Further bending techniques and crystal materials can be investigated,
including for instance low-Z (diamond)
and high-Z (germanium, tungsten) \cite{cern,ge},
as options for scraping facilities.

\end{document}